# Quantum field theory solves the problem of the collapse of the wave function

Alexey V. Melkikh,

Ural Federal University, Mira street 19, Yekaterinburg 620002, Russia

The problem of measurement in quantum mechanics is that the quantum particle in the course of evolution, as described by the linear Schrodinger equation, exists in all of its possible states, but in measuring, the particle is always detected in only one of its states. This property is called the "collapse of the wave function" and was formulated by Von Neumann as one of the postulates of quantum mechanics. However, it remains unclear at what point in time and under what laws this transition occurs. This article demonstrates that the collapse of the wave function may be due to the creation or annihilation of particles (quasi-particles). The processes of the creation or annihilation of particles play a key role in the measurements and are described on the basis of quantum field theory. The system of equations of quantum field theory of particles and fields is non-linear; as a result, the principle of superposition does not hold for the theory. The collapse of the wave function is a consequence of this non-linearity and occurs at the moment of creation (annihilation) of a particle. This result demonstrates that the wave function collapse can occur in both microscopic and macroscopic systems. Understanding the mechanisms of the collapse of the wave function can lead to the creation of microscopic devices involved in the calculations based on quantum computing.

The problem of performing measurements in quantum mechanics has again become relevant in recent years (see, for example [1-3]), largely because of the quantum technologies that are available to work with individual particles. The problem is that the principle of superposition in quantum mechanics contradicts one of its basic postulates: von Neumann's postulate of the collapse of the wave function. The Copenhagen interpretation of quantum mechanics suggests that the collapse of the wave function is a consequence of the interaction of the particle with a macroscopic measuring device, which is a classical object and cannot simultaneously be in multiple states [4, 5]. However, it remains unclear what the difference is between the device and an arbitrary quantum system, which must obey the principle of superposition. At what point of time should be a transition from the motion of a particle described by the linear Schrodinger equation to the collapse of the wave function? There are a number of other interpretations of quantum mechanics that provide different explanations of the wave function collapse (see, e.g. [6-8]). For example, according to [9], all approaches to the problem of the collapse of the wave function can be divided into three groups:

- do not modify quantum mechanics but modify its interpretation (Everett),
- do not modify quantum mechanics but modify its mathematical formulation (Bohm),
- replace the current quantum theory with a more general theory, which, in the respective limits, corresponds to the quantum and the classical mechanics [6, 9, 10].

Note that none of these directions has been supported by experimental evidence.

Thus, one of the most important problems of quantum mechanics - the problem of the collapse of the wave function (the problem of measurement) - remains unresolved.

There is, however, another possibility to explain the causes of the collapse of the wave function (solution to the problem of measurement), which is not associated with any additional assumptions or the creation of a new theory.

One of the important properties of the measuring devices is that a qualitative change occurs in them, which, to a certain extent, adapts to the properties of measured system. If we consider such qualitative changes at the microscopic level, we can see that the simplest of these processes is the creation or annihilation of particles (quasi-particles). Indeed, any measurement process is accompanied either by the creation or annihilation of particles. For example, to fix the position of the electron on the screen, it is necessary that the electron knocks out at least one photon or another electron. These secondary particles, in turn, lead to the creation of other particles, thereby amplifying the resulting signal to a macroscopic level, which we can observe using instrumentation. Of course, the collapse of the wave function does not need to be associated with the term "device" or any technological process but must occur in any macrosystem.

The basic assumption with which the problem of the collapse of the wave function can be solved is that the birth (annihilation) of the particles plays a key role in this process.

In addressing the problem of measurement in quantum mechanics, the mechanics of particles is often observed as a closed science, and mechanics of fields is not usually taken into account. However, because the particles and the fields are connected with each other, the closure can be attributed only for particles and fields together. Of course, there are processes for which the Schrödinger equation provides an ample description. In this case, an account of the quantization of fields can be performed phenomenologically, bearing in mind that, for example, the energy of the emitted photons corresponds to the difference between the energy levels of electrons in the atom. However, there are processes for which the quantization of fields is important. These processes include, in particular, the measurement process. Let us consider the processes of creation and annihilation of particles on the basis of quantum field theory.

For example, in quantum electrodynamics (QED), the Dirac field and the electromagnetic field are operators (see, eg. [11]). These operators satisfy the coupled system of equations of motion of QED. These equations can be written as follows:

$$\left\{\gamma_\mu\left(\frac{\partial}{\partial x_\mu}-ieA_\mu(x)\right)+m\right\}\psi(x)=0, \quad (1)$$

$$\left\{\gamma_\mu^T\left(\frac{\partial}{\partial x_\mu}+ieA_\mu(x)\right)-m\right\}\bar{\psi}(x)=0, \quad (2)$$

$$\Box A_\mu(x)=-j_\mu(x), \quad (3)$$

where, $\psi(x)$ is the wave function, and $j_\mu(x)$ is the 4-density of the electron current, which is equal to

$$j_\mu(x)=\frac{ie}{2}\left(\bar{\psi}(x)\gamma_\mu\psi(x)-\bar{\psi}^c(x)\gamma_\mu\psi^c(x)\right),$$

where the subscript "c" denotes the change of the charge sign, $A_\mu$ is the potential of the electromagnetic field, $\gamma_\mu$ are the Dirac matrices, and $e$ is the electron charge,

$$\Box\equiv\frac{\partial^2}{\partial t^2}-\frac{\partial^2}{\partial x^2}-\frac{\partial^2}{\partial y^2}-\frac{\partial^2}{\partial z^2}.$$

Because this system of equations of motion do not permit an exact solution, it is solved approximately by the method of perturbation theory on the existing small dimensionless parameter α = 1/137 (the fine-structure constant). An important property of such a system is its *non-linearity*.

In many experiments, the most important process is the scattering by some particles on the others. In such a situation, it suffices to know the asymptotic behavior of the particles after the collision, which is described by the S-matrix. According to quantum theory, the amplitude of a physical process is expressed through the matrix element of the *S*-matrix. In turn, the *S*-matrix is expressed through the *T*-exponent of the interaction Lagrangian *L*:

$$M = \langle out|S|in\rangle, \quad S = T\exp\left(i\int dx L_I(\varphi(x))\right).$$

Interaction Lagrangians are constructed based on the quantum fields φ, and the operation T puts these fields in order of the time argument $x_0$. In this case, the exponent is understood in the sense of the series, each member of which corresponds to the creation (or annihilation) of the particles and can be described using the corresponding Feynman diagrams. Note that, in this case, the dependence of the series members on the wave functions in S-matrix decomposition is also essentially nonlinear.

Thus, the principle of superposition for the system "particles + fields" is not satisfied (i.e., the sum of the solutions is not the solution). As is known, the superposition principle in quantum mechanics is the main obstacle to the understanding of the mechanism of wave function collapse. The failure of superposition principle, consequently, involves the collapse of the wave function. In this case, the collapse that is a discontinuous (discrete) event naturally corresponds to a different discrete event - the creation (annihilation) of the particle. At the moment of the creation (annihilation) of a particle, the Schrödinger equation ceases to describe the evolution of the system. At this moment, there is a transition from the probability amplitudes for the probabilities themselves of finding the particle in some region of space under the Born rule. For example, if the device is initially in a superposition of its two states ($|\Phi_1\rangle, |\Phi_2\rangle$), and the particle is in a superposition of its two states ($|\psi_1\rangle, |\psi_2\rangle$), then the result of the interaction of the complete system must be described by a wave function:

$$|\Psi\rangle = c_1 |\psi_1\rangle |\Phi_1\rangle + c_2 |\psi_2\rangle |\Phi_2\rangle.$$

However, as a result of the creation (annihilation) of the new particle, a transition from one of two states occurs

$$|\Psi_2\rangle = |\psi_2\rangle |\Phi_2\rangle, \quad \text{or} \quad |\Psi_1\rangle = |\psi_1\rangle |\Phi_1\rangle.$$

According to the Born rule, the probability of finding the particle in any state is proportional to the square of the wave function

$$P = |\Psi|^2.$$

At first glance, QED in many cases gives only a slightly more accurate picture than quantum mechanics. Indeed, for example, the calculations of the energy levels of the hydrogen atom by quantum mechanics result in very accurate values. QED still refines them, but by only a small amount, which in many cases is not important (e.g., the Lamb shift of energy levels). Why, in this case, can we work without a complete set of equations (1-3) and solve the linear Schrodinger equation? This solution can be achieved because the spectral lines are measured by macroscopic devices (for example, the registration of the spectrum of the hydrogen atom with a spectrometer is a macroscopic process) in which the wave function collapse is inevitable, because of a large number of produced particles. That is, the non-linearity introduced by QED and the failure of the superposition principle act as if they were hidden inside a macroscopic device. When constructing a model of the atom, we do not consider the measurement device and how the device works; instead, we consider by default the device as being macroscopic.

Thus, the evolution of a quantum particle in a medium can be summarized as follows: while the (quasi)particles are not created and annihilated the motion of the particle is coherent (the particle is simultaneously in all of its possible states); as soon as any particle is created (annihilated) as a result of the interaction with the environment, the collapse of the wave function occurs (from the superposition of its states the particle transfers to one of them); subsequently, coherent evolution continues from the collapsed wave function state.

The non-linearity of the equations of quantum field theory is also distinctive for quasi-particles, such as phonons and magnons. For example, the scattering of electrons on atoms of a solid lattice can be both elastic and inelastic. As a result of inelastic scattering, one or more phonons can be created, and this process is essentially non-linear. The created phonons can be used to obtain information about the electron. In the case of elastic scattering, when phonons are not created, additional information about the electron cannot be obtained. Mathematically, this conclusion is expressed in the fact that the von Neumann entropy (a generalization of the Shannon entropy for quantum processes) of a pure (coherent) quantum state is equal to zero:

$$S(\rho) = -Tr\hat{\rho}\log_2 \hat{\rho},$$

where $\hat{\rho}$ – is the density matrix.

As a result of the measurement, the state becomes mixed, and the entropy becomes positive. If the state remains pure, the von Neumann entropy is equal to zero; hence, new information about the system cannot be obtained. In any case, receiving information requires the collapse of the wave function.

If we describe the measurement process by Feynman diagrams, we can see that the measurement (in which the collapse of the wave function occurs) corresponds to the diagrams in which real particles are created (annihilated). All other processes correspond to the diagrams in which real particles are not created, and only virtual particles participate, wherein the collapse of the wave function does not occur.

Consider, for example, the classic experiment on the interference of electrons incident onto two slits. In this experiment, two phases can be distinguished: the electron motion in space and an electron registration on the screen. If, during the interaction of electrons with the matter of the diaphragm, no new particles (e.g., phonons) are created (and they do not disappear), there will be an observed interference picture on the screen (Fig. 1).

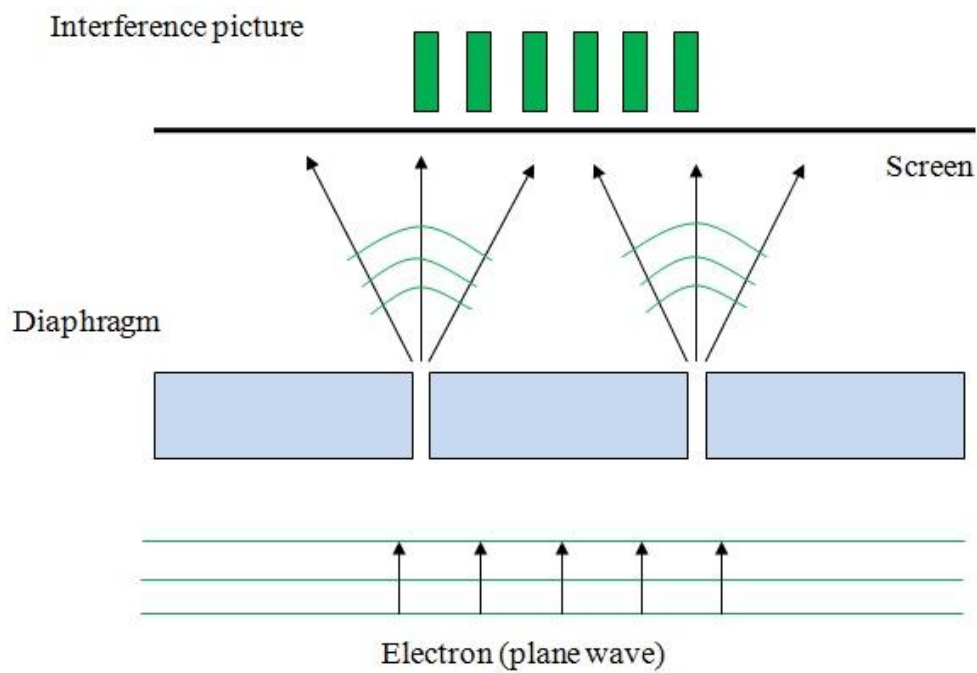

Fig. 1 For elastic electron scattering by the particles of the medium, an interference pattern is observed on the screen.

Observing such an interference pattern may result in the measurement of the position of the particles on the screen. The wave functions of the measured particles then collapse. If, as a result of the interaction of the electron with the material of the diaphragm creates at least one particle (e.g., a phonon), then there was an intermediate position measurement of an electron (thus making it possible to determine through which slit the electron passed). In this case, the picture on the screen (which can also be obtained only as a result of measurement) will change:

if, during repeated measurements, phonon production occurs at different points in space, the interference pattern will disappear (Fig. 2), and two peaks are observed in front of each slit.

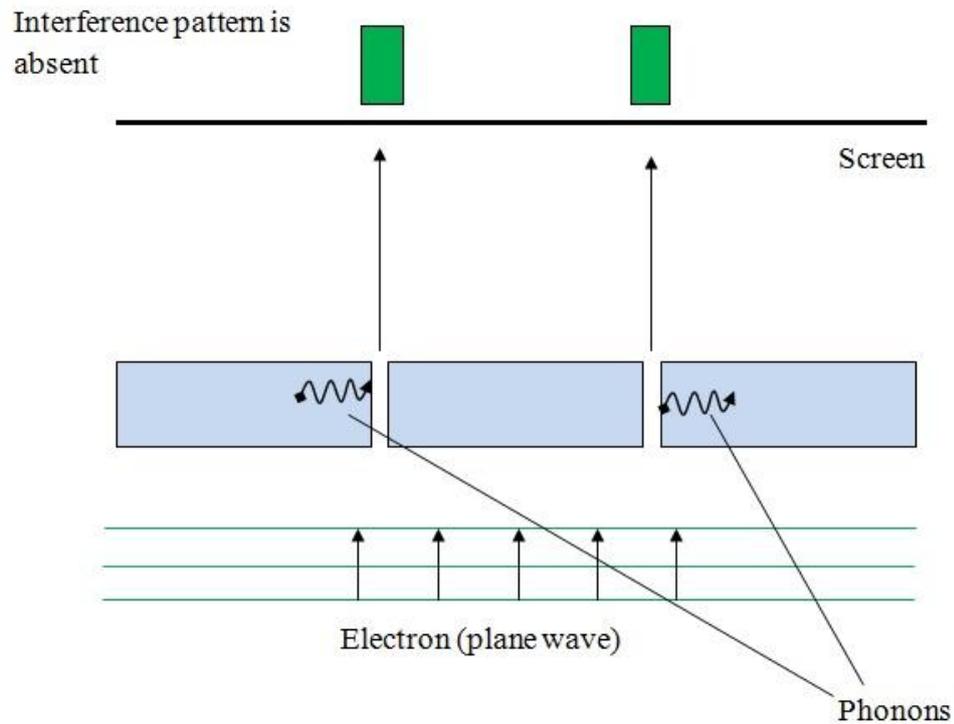

Fig. 2 For inelastic electron scattering of particles of the medium in which the phonons are created (disappear), the interference pattern on the screen does not appear.

Thus, phonons act as microscopic devices: they already carry information about the electron properties at their birth. Naturally, this signal may be further enhanced by the creation of particles.

This applies, of course, not only to the interaction of electrons with phonons but also to other particles and quasi-particles.

Thus, all quantum events can be divided into two classes:

- The movement of particles in space, accompanied by their elastic scattering by the particles of the medium,

- Inelastic scattering by the particles of the medium, accompanied by creation (annihilation) of other particles.

In the first case, a collapse of the wave functions does not occur. In the second case, the collapse of the wave function occurs, which is an inherent part of measuring devices. Only the second process allows us to receive information about the original system.

Thus, the solution to the problem of measurement in quantum mechanics is to recognize that the system of "particles + fields" has a natural mechanism of decoherence associated with the creation (annihilation) of particles in the quantum field. Regarding the system as a whole, the superposition principle does not hold. That implementation of the principle of superposition to the system as a whole leads to a paradoxical of measurement process and the collapse of the wave function.

After the initial registration connected with the particles creation, the effect should increase for registration on a macroscopic device. During this strengthening, a further creation of particles (quasi-particles) occurs. This follow-up process is not fundamentally different from the initial measurement.

In a number of articles [12, 13], the so-called "interaction-free measurements" was considered. According to the corresponding authors, these interaction-free measurements enable the determination that the relevant body is in a certain area, without interacting with the body. This effect is essentially quantum, and can be used according to the authors for the detection of the fragile object state.

What can be said about such an effect in terms of the above-proposed concept of measurement associated with the creation of (quasi)particles?

First, the accuracy is necessary to understand the term "interaction-free measurement" (this was also mentioned in the article of Vaidman [13]). If the interaction of the body with a photon did not exist, the photon would not play any role in determining whether the body is present in this place; in such a case, it is impossible to determine the location of the body. However, if we consider the creation of the particles as a key aspect of the interaction in relation to the collapse of the wave function, it is easy to understand that the interaction-free measurements is connected with absence of real particles in the experiment. This interaction-free measurement process involves only virtual particles.

Second, can we call this interaction a "measurement"? The answer is "no", because the measurement in this experiment (and in any other) occurs only as a result of the collapse of the wave function through the detection of photons (or other particles) by detectors. "Interaction-free measurements" are a special case of a wide class of phenomena of quantum interference and diffraction. Indeed, when the observation of the interference pattern, such as from a disc, we can obtain information not only about the presence of the disk but also about its shape. In the observation of electron interference by two slits, one can determine, for instance, that exactly two open slits exist. In such a case, any motion of a quantum particle is described by the Schrödinger equation (as long as the measurement has not yet occurred), and has the same property as the "interaction-free measurement": it provides information about the environment in which this motion occurs. However, this information can only be obtained as a result of a "true" measurement in devices, in which the particles are created and, as a consequence, a collapse of the wave function of the measured particles occurs. Thus, the term "interaction-free measurements" seems superfluous because it can be related to any of quantum processes.

Note also that in «weak measurements» [14, 15] and «non-demolition measurements» [16-18] collapse of the wave function takes place in any case. Only as a result of the collapse of the wave function can the information about the measured system be obtained.

One of the fundamental issues related to the concept of wave function collapse is the question of why macroscopic objects are not in a state of superposition (see, for example [9]).

Based on the above assumptions, we can give the following answer to this question: macrosystem (for example, Schrödinger's cat) is not in a superposition of its states because the larger the system, the more likely is the creation or annihilation of particles in it (for example, a black body constantly emits and absorbs photons). This means that the time of existence of a macroscopic system in a pure state will be exponentially small (except for special cases, such as superfluidity and superconductivity, when the particle emission is forbidden by the laws of quantum mechanics).

We can also answer the question of whether there is some general dynamics that exists in the limits of both classical and quantum mechanics? Such dynamics exist - it is a joint quantum theory of fields and particles. In one limit, the joint quantum theory gives the quantum mechanics of particles (for problems in which the quantization of fields inessential), and in the other limit (large bodies), when the particles are created and annihilated in large numbers, the bodies move classically because their wave functions almost continuously undergo collapse. This means that we can consider their trajectories and other classical properties.

If we consider the interpretations of quantum mechanics, they are often considered only as a supplement to the philosophy of quantum theory. In many cases, ("for all practical purposes") it is irrelevant what interpretation of the quantum process we choose because the mathematics of quantum mechanics gives a very precise answer. However, when considering the problem of measurement, this is not so.

Why does the postulate of von Neumann work so well, even though it is not a question at all about the creation (annihilation) of particles? This is because the vast majority our instruments are macroscopic. This macroscopic nature of the instruments means that the creation of particles during their operation occurs with overwhelming probability as an effect of signal amplification.

However, for microscopic devices, wave function collapse may not occur. In this case, the postulate of von Neumann does not apply.

Nanotechnologies are currently being developed that can perform measurements without human participation. In this case, the measuring devices need not adapt to our senses (such as a voltmeter and thermometer). These devices must adapt to the specific size and characteristic times of the technology. In this sense, the signal strengthening to macroscopic values may not be necessary. The device itself in this case will be essentially microscopic and a quantum object.

The claim that the collapse of the wave function is a consequence of the creation (annihilation) of particles can be tested experimentally. Consider, for example, the interference of an electron on two slits (Fig. 3).

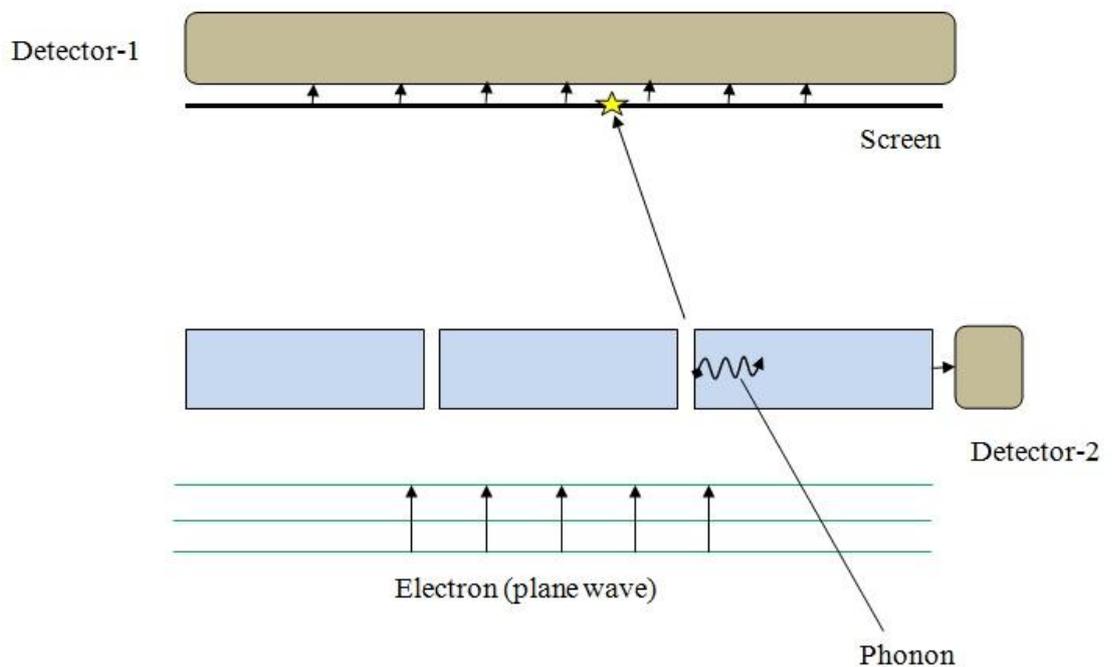

Fig. 3 Statistical relationship between the incidence of the electron in the interference peak and the creation of a phonon can be obtained experimentally.

If a substance that interacts with the electron represents a microscopic crystal at a low temperature, there will be a relatively small number of phonons in it. If a single detector is used to register of an electron at a particular location on the screen, and another detector is used to register the creation of a phonon (vibration of the crystal), it is possible to obtain information about the relationship between these processes. We assume that both of the detectors are macroscopic, and the wave function collapse in them inevitably occurs.

The outcome of the measurements will be substantially different, depending on whether at least one phonon created during interaction with the electron and the crystal. If a phonon was created, the electron will be registered on the screen at a certain place, which does not comply with (on average) the positions of the interference maxima. At the same time, the second detector should function to register a phonon. The registration result of a phonon can be further strengthened, which, in the end, will be macroscopic (e.g., it may be a change in the computer's memory, which stores the result of the measurement).

If there is only elastic scattering of the electron (i.e., a phonon was not created) the detector-1 is more likely to register an electron on the screen at positions of the interference maximum (because an interference pattern is observed only in the absence of decoherence) and the second detector does not detect a signal. Naturally, repeated measurements of the electron will be positioned at different locations on the screen, but an examination of the distribution of a set of measurement results will clearly indicate a correlation between the creation of the particle and the wave function collapse. For example, the distance between the position of the electron to the interference maximum and the triggering of a second detector that registers the phonons can serve as a matching criterion for correlation.

In this experiment, the microscopic crystal with the created phonon acts as a microscopic device, and the sensor detecting the presence of a phonon acts as a macroscopic device that records the state of the microscopic device. If the correlation is found, it can be concluded that the collapse of the wave function of an electron is really a consequence of the creation of a phonon. Having performed similar experiments with particles of different types, conclusions about the causes of the collapse of the wave function can be generalized for a wide class of quantum processes. Naturally, the registration of a single particle often represents a significant amount of technical difficulties. However, these difficulties are not of a fundamental nature.

Thus, if one is to associate the concept of "measurement" with the creation (annihilation) of particles (quasi-particles), the collapse of the wave function receives its natural explanation. Solving the problem of measurement in quantum mechanics is associated with a substantial non-linearity of the system of equations describing the particles and fields, for which the superposition principle does not hold. In this case, there is no need to consider the device as macroscopic - a collapse will also occur for microsystems. Macroscopic bodies are not in a state of superposition because within them, the creation and annihilation of particles constantly occurs, thereby destroying coherence. Changing the number of particles in the process is a prerequisite for the collapse of the wave function. Is this change a sufficient condition? This question should be clarified by further experiments.